# Tunneling Time in Ultrafast Science is Real and Probabilistic

**Authors:** Alexandra Landsman[1*], Matthias Weger[*], Jochen Maurer, Robert Boge, André Ludwig, Sebastian Heuser, Claudio Cirelli, Lukas Gallmann, Ursula Keller

**Affiliation:** Physics Department, ETH Zurich, CH-8093, Zurich, Switzerland

[1]To whom correspondence should be addressed: alexandra.landsman@phys.ethz.ch

[*]These authors have contributed equally to the work

**Abstract**: We compare the main competing theories of tunneling time against experimental measurements using the attoclock in strong laser field ionization of helium atoms. Refined attoclock measurements reveal a real and not instantaneous tunneling delay time over a large intensity regime, using two different experimental apparatus. Only two of the theoretical predictions are compatible within our experimental error: the Larmor time, and the probability distribution of tunneling times constructed using a Feynman Path Integral (FPI) formulation. The latter better matches the observed qualitative change in tunneling time over a wide intensity range, and predicts a broad tunneling time distribution with a long tail. The implication of such a probability distribution of tunneling times, as opposed to a distinct tunneling time, challenges how valence electron dynamics are currently reconstructed in attosecond science. It means that one must account for a significant uncertainty as to when the hole dynamics begin to evolve.

**Main Text:** The question of how long a tunneling particle spends inside the barrier region has remained unresolved since the early days of quantum mechanics [1]. The main theoretical contenders, such as the Buttiker-Landauer [2], the Eisenbud-Wigner (also known as Wigner-Smith) [3], and the Larmor time [4, 5] give contradictory answers . Recent attempts at reconstructing valence electron dynamics in atoms and molecules have entered a regime where the tunneling time genuinely matters [6-10]. Nevertheless, common reactions in the broader scientific community to the tunneling time problem can be roughly grouped into two categories: 1) "it's easy" or 2) "it can't be done".

In the first category, it is sometimes suggested that the tunneling time is instantaneous because both the Keldysh [11] and the closely related Buttiker-Landauer [2] times are imaginary (corresponding to the decay of the wavefunction under the barrier). However, both of these times were obtained using physical considerations of what a tunneling electron actually "sees" in real time while inside the barrier [2,11]. To cite Landauer: "More important than the exact results and its relation to theoretical controversies, is the fact that a timescale associated with the barrier traversal can be measured, and is a real (not imaginary) quantity." [12].

In the "it can't be done" category, it is often said that quantum mechanical uncertainty does not allow for a deterministic tunneling time. However, the main theoretical contenders for tunneling time should be viewed as average values, rather than deterministic quantities [13]. In fact, while the four well-known times considered here were obtained using very different physical models, they can also be expressed as the outcome of an averaging procedure using the tunneling time probability amplitude, obtained via the Feynman path integral (FPI) approach [13, 14]. The FPI approach is a natural way to extract tunneling time, since each path is deterministic and therefore has a definite time associated with it. The resulting tunneling time probability amplitudes are however complex and interfere. As pointed out in [1], while no one disputes the accuracy of this construction, it is not clear what procedure to use for calculating relevant physical quantities with FPIs.

Here, we turn to experiment to resolve this issue and shed light on how tunneling times are to be calculated for real-world applications. Prior experiments were either not sufficiently precise or did not explore a sufficient range of barrier shapes to allow for a detailed comparison with theory. To date only two experiments [15, 16] were done at the single particle level, thereby avoiding the possibility of a "pulse reshaping" process, which can lead even to negative tunneling times [16, 17]. Prior attoclock measurements [15] found an upper limit on tunneling time of around 40 as, but within a narrow intensity range of $3.3-4.8\times10^{14}\,W/cm^2$, and hence observed no significant trend in tunneling time. More recent attoclock measurements in helium and argon extended towards higher intensities also did not resolve any tunneling delay times [18]. Another recent experiment measured the time an electron involved in high harmonic generation (HHG) exits the barrier [19]. However, the absolute timing of ionization and therefore the tunneling delay time could not be extracted [20].

Of the five theoretical approaches considered, two cannot be excluded by our measurements: the Larmor time and the probability distribution of tunneling times constructed using FPIs. These two are not mutually exclusive. In particular, the correctness of Larmor time does not preclude the existence of a probability distribution of tunneling times. The theoretically determined distribution that is compatible with our experimental results is broad at the relevant time scales and has a rather long tail, suggesting that assuming a single tunneling time (instantaneous or otherwise) is potentially a deep flaw in ultrafast science applications.

**Experiment:** The electric field of an intense laser pulse bends the binding potential of an atom, creating a barrier, whereby electrons can tunnel into the continuum, be accelerated by the laser field, and eventually register their momenta at the detector. The momentum distributions were measured by two different experiments: a cold-target recoil-ion momentum spectrometer (COLTRIMS) [21] and a velocity map imaging spectrometer (VMIS) [22]. At lower intensities, attoclock [15, 23] measurements were performed with VMIS with the gas nozzle integrated into the repeller plate (see Fig. 1c). The integration of the gas nozzle allows one to achieve target gas densities that are significantly higher compared to setups employing cold atomic beams [24]. Two-dimensional projections of the photoelectron momentum distribution were recorded in steps of two degrees covering 180 degrees. The three-dimensional momenta distribution, and thus the electron momenta distribution in the polarization plane, was retrieved by tomographic reconstruction, shown in Fig. 1d (more detail in S.1). The COLTRIMS measurements were performed in a similar way as described elsewhere [18]. Figure 2 shows good overlap between COLTRIMS and VMIS data, with error bars higher on COLTRIMS.

The experiment spans a wide intensity range of $0.73-7.5\times10^{14}\,W/cm^2$, corresponding to variation in the barrier width by about a factor of 3 in the range from 7.5 to 24 au, and given approximately by $I_p/F$, where $I_p$ is the ionization potential and $F$ the peak electric field. An important parameter in strong field ionization is the Keldysh $\gamma = \omega\sqrt{2I_p}/F$ [11], which divides the "vertical channel" of multi-photon ionization ($\gamma \gg 1$) from the "horizontal channel" of optical tunneling ($\gamma \ll 1$). The experimental regime is in the $0.8 < \gamma < 2.5$ range, corresponding to "nonadiabatic tunnel ionization" [25, 26]. In this regime, while the tunneling probability may be substantially modified from the quasistatic rates given in [11, 27], phase-independent contributions due to multiphoton absorption are small (approximately 3.3% of the total rate for $\gamma \sim 2$ [25]), and tunneling remains the dominant ionization mechanism, widely used to investigate molecular orbitals [8, 28, 29] and electron rearrangement [6-10] after ionization.

To map the measured momenta of the electrons, shown in Fig. 1, to the phase of the electric field, $\theta_i$, at which the electron first appears at the tunnel exit, we use attosecond angular streaking [23]. The electron located at the peak of the momenta distribution, given by $\left|\langle \vec{k}|\Psi(t)\rangle\right|^2$, corresponds to the most probable electron trajectory [26]. To locate this peak from measurements, radial integration is used, combined with an asymmetric Gaussian fit to extract the angle at which the maxima in the distribution occurs (see Fig. 1a,b), corresponding to ionization at the peak of the laser field. This measured angle, $\theta_m$, shown in Fig. 2, is used to extract tunneling time after subtracting the Coulomb correction, $\theta_{Coul}$, and the streaking angle, $\theta_{str}$, which includes rotation due to the drift created by the vector potential of the electric field, resulting in the experimentally measured tunneling time, $\tau$, given by: $\omega\tau = \theta_i = \theta_m - \theta_{Coul} - \theta_{str}$, where $\omega$ is the central frequency of the laser. In calculating $\theta_i$, non-adiabatic effects and an offset of the streaking angle from 90 degrees were taken into account, resulting in a minor (less than 13 as) correction to tunneling time (see S.2).

**Theory**: For direct comparison with the experiment, analytic calculations are done for the Fourier component, $\vec{k}$, of the electron wavepacket that maximizes $\left|\langle \vec{k}|\Psi(t)\rangle\right|^2$, corresponding to the peak of the electron momenta distribution, from which the tunneling time is experimentally extracted, as described above. The tunneling process acts as a momentum filter that maps different Fourier components of the bound-state wavefunction onto different momenta, $\vec{k}$, at the tunnel exit with probability $\propto \exp(-k_\perp^2/2\sigma_\perp^2)$ [28], where $k_\perp$ is the momentum transverse to the direction of tunneling and $\sigma_\perp$ is given by ADK [27].

The maximum of $\left|\langle \vec{k}|\Psi(t)\rangle\right|^2$ is given by the transmission of a $\Phi(x,\vec{k}_\perp=0)=\Phi(x)$ component, in the partial Fourier transform [30] of the bound-state wavefunction:

$\Psi_b(x,y,z) = \frac{1}{2\pi}\int dk_y \int dk_z e^{i\vec{r}\cdot\vec{k}_\perp}\Phi(x,\vec{k}_\perp)$, where x is the major axis of polarization. The tunneling times were calculated for this component $\Phi(x)$ within the short-range potential approximation, taking into account non-adiabatic effects. The robustness of the results to specific barrier shape and non-adiabatic effects are discussed in S.3. Sensitivity to barrier shape was tested using a square barrier of the same width, which resulted in less than 50% variation in tunneling time. Since the actual barrier shape is much closer to the triangular shape than the square barrier, we expect the deviation due to the short-range potential approximation to be well below 50%. The contribution of non-adiabatic effects to tunneling time estimates is small, even in the case when the ionization rates are substantially offset from the quasi-static. This is partly due to the exponential dependence of ionization probability on barrier width, but only a linear dependence of tunneling time (see Fig. 3d).

The four widely used tunneling times were calculated by finding the transmission amplitude for the propagation of $\Phi(x)$ through the potential barrier, given by: $T = |T|e^{i\phi}$, and using the following definitions:
$\tau_{BL} = -\hbar \partial \ln|T|/\partial V$; $\tau_{LM} = -\hbar \partial \phi/\partial V$; $\tau_{PM} = \hbar \partial \ln|T|/\partial E$; $\tau_{EW} = \hbar \partial \phi/\partial E + w/k$, for the Buttiker-Landauer [2, 14], Larmor [4, 5], Pollack-Miller [31] and Eisenbud-Wigner times [3], respectively (here $V$ is the height of the barrier, and $E$ is the electron energy). An additional term, $w/k$, is present in $\tau_{EW}$, where $w$ and $k$ are the barrier width and electron velocity, respectively. This additional term corresponds to the propagation of the electron in the barrier region if that barrier were absent, and has to be added to get the total time [17], since the first term only gives a *relative* time shift [3]. The Eisenbud-Wigner time has been used extensively to explain the relative single photon ionization delay between ionization of a 2s and a 2p orbital in Neon, observed in [32]. Perhaps counter-intuitively, it is not the actual time it takes to absorb a photon (which cannot be calculated using this definition), but rather an additional phase shift in the peak of the propagating wavepacket induced by the presence of the ionic potential *after* that photon is absorbed. Calculating the shift in the peak of the wavepacket is straight-forward in single photon ionization, where the total energy of the electron is above threshold, leading to a well-defined single peak and propagation of the entire wavepacket. However, the Eisenbud-Wigner time is much more disputed in tunneling [1], where the peak of a wavepacket is absent inside the barrier [17], and moreover, a large part of the wavefunction remains confined.

The predictions of the four tunneling times are shown in Figure 3a. Although these times were derived using very different physical models, they have been shown to arise from various averaging procedures using the tunneling time probability amplitude, $f(\tau)$, constructed with FPIs [13, 14, 33, 34]. In particular, the Buttiker-Landauer and the Larmor times correspond to the absolute and the real parts, respectively, of the following complex-valued average [1, 13, 33]: $\langle \tau \rangle = \int_0^\infty \tau f(\tau) d\tau / T$. While, as mentioned in the introduction, it is not clear what averaging procedure to use with $f(\tau)$, the above definition has been widely used, in part because it can be expressed as the transition element [34]: $\langle \tau \rangle = \langle \Psi_i | \tau | \Psi_T \rangle$, where $\Psi_i$ and $\Psi_T$ are the normalized incident and transmitted parts of the wavefunction, respectively. Perhaps surprisingly, $|\langle \Psi_i | \tau | \Psi_T \rangle|$ is far too large, and only the real part of $\langle \tau \rangle$, equivalent to $\tau_{LM}$, is within our experimental uncertainty.

The FPI approach is particularly appealing, because the total transmitted wavefunction can be expressed as a sum of all possible paths, each path corresponding to a deterministic tunneling time and contributing $\exp(iS[x(t)]/\hbar)$. The quantity $f(\tau)$ represents the contribution to the total transmission amplitude of only those paths that spend an amount of time, $\tau$, inside the barrier, such that: $T = \int_0^\infty f(\tau) d\tau$.

Using $f(\tau)$, we construct the probability distribution of tunneling times, following the method in [35] (see also S.3), and shown in Fig. 4. The peak of this probability distribution is shown along with other theoretical predictions in Fig. 3a,b. This peak corresponds better to the experimental observable (which is the peak of the recorded electron momenta distribution), than the expectation value, given by the other tunneling time definitions. This is because trajectories that begin to tunnel at the peak but have longer or shorter tunneling times than the most probable trajectory will not end up at the peak of the momenta distribution, but will nevertheless be included in any averaging procedure that extracts the expected value of tunneling time.

Historically, measured tunneling time varied with the nature of the experiment. Besides the attoclock measurements, only one other experiment was done at the single particle level [16], thereby avoiding the possibility of a pulse reshaping process [1]. Our findings are consistent with this single photon experiment [16], in that a particle moving through a potential barrier takes significantly less time than free propagation over the same distance. Therein this yielded superluminal velocities[16]; we get tunneling times that are just a small fraction of the free propagation time (though not superluminal, as Fig. 3d shows) for electrons with kinetic energies characteristic of electron motion in Helium.

As Fig. 3 shows, of the five theoretical approaches, two cannot be excluded: the Larmor time and the probability distribution of tunneling times constructed using FPIs. These two are not mutually exclusive. In particular, the correctness of Larmor time (viewed as an average, rather than a deterministic, quantity) allows the existence of a probability distribution of tunneling times. The probability distribution shown in Fig. 4 has a long asymmetric tail that lengthens, along with an increase in the position of the peak and the full-width-half-maxima (FWHM), as intensity decreases: this suggests that both uncertainty and expected duration of tunneling time increase at lower intensities, corresponding to a larger barrier width.

Our results show that the probability distribution of tunneling time, at all measured intensities, adds significant uncertainty to reconstruction of attosecond electron dynamics after strong field ionization. In particular, the FWHM is larger or comparable, depending on intensity, to the "universal attosecond response to removal of an electron" of about 50 *as*, found computationally [36]. Hence, the standard assumption, of electron hole dynamics beginning to evolve at the instant the electron appears at the tunnel exit [7, 8, 36], is highly problematic, since it is likely that by that time, important processes associated with the evolution of a hole in any atom or molecule have already taken place. A second crucial issue is the loss of coherence of the hole if tunneling time is probabilistic, rather than deterministic. This loss will depend on the time scale of the evolution of a hole [37] and is therefore likely to be much more significant for molecules than for atoms. This is because the hole period is determined by the energy splitting of nearby valence orbitals, resulting in a period of 6.2 fs in Krypton [37, 38], but only around 1.2 fs in $CO_2$ [8]. As Fig. 4 shows, the long tail of the distribution can extend to a substantial fraction of a hole period, resulting in a substantial loss of coherence.

The time-scale of tunneling, once an unresolvable question for theorists of the foundations of quantum mechanics, therefore meaningfully affects the reconstruction of electron dynamics using HHG [8] or pump-probe experiments [7] – which is the primary goal of ultrafast science. The implications for both ultrafast experiment and theory, especially for molecules and at time resolutions of current and future interest [39], are at once fundamental, practical, and approachable with existing technology.

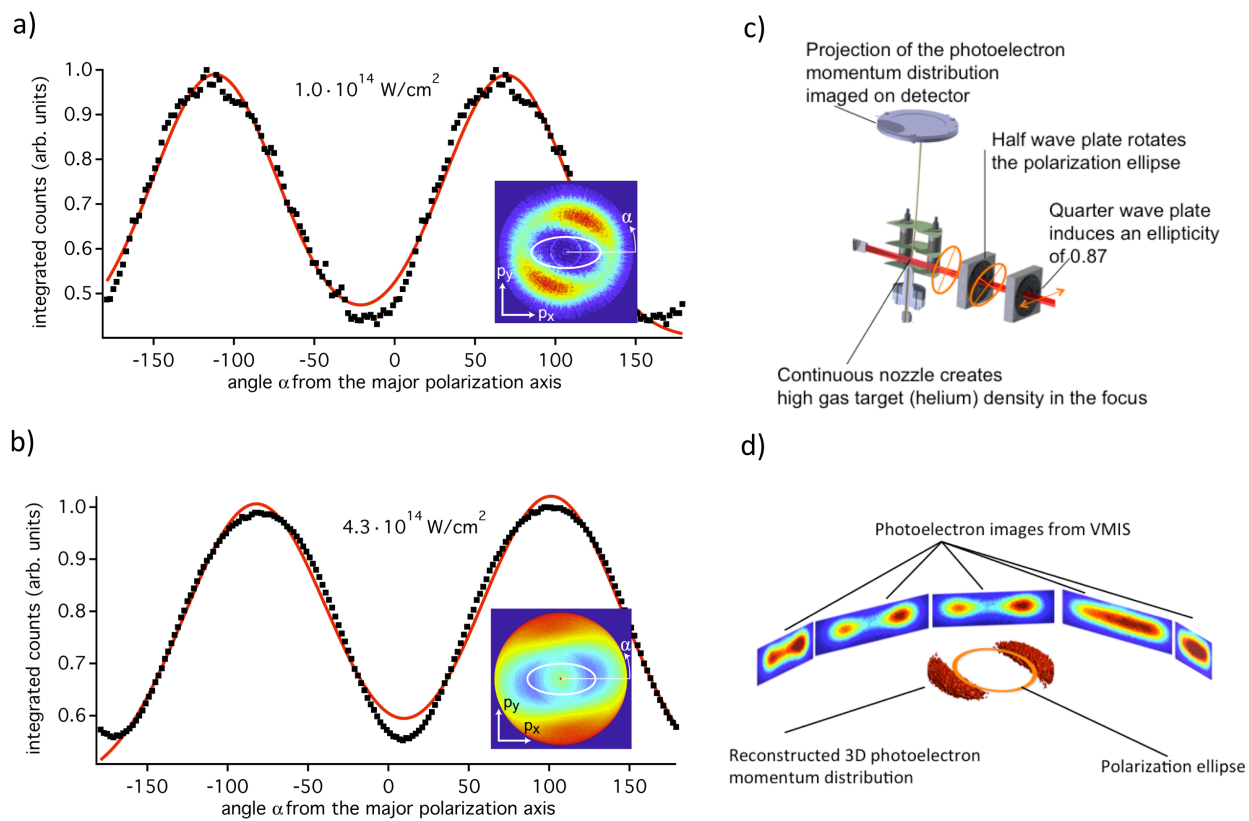

Fig. 1. Measurement of electron momenta distributions using velocity map imaging spectrometer (VMIS). a) Fitting of peaks in electron momenta distribution with a Gaussian at lower intensity and clockwise polarization. b) Higher intensity and counter-clockwise polarization. c) Experimental set-up. d) Reconstruction of full 3-D electron momenta distribution using projections in the momentum space.

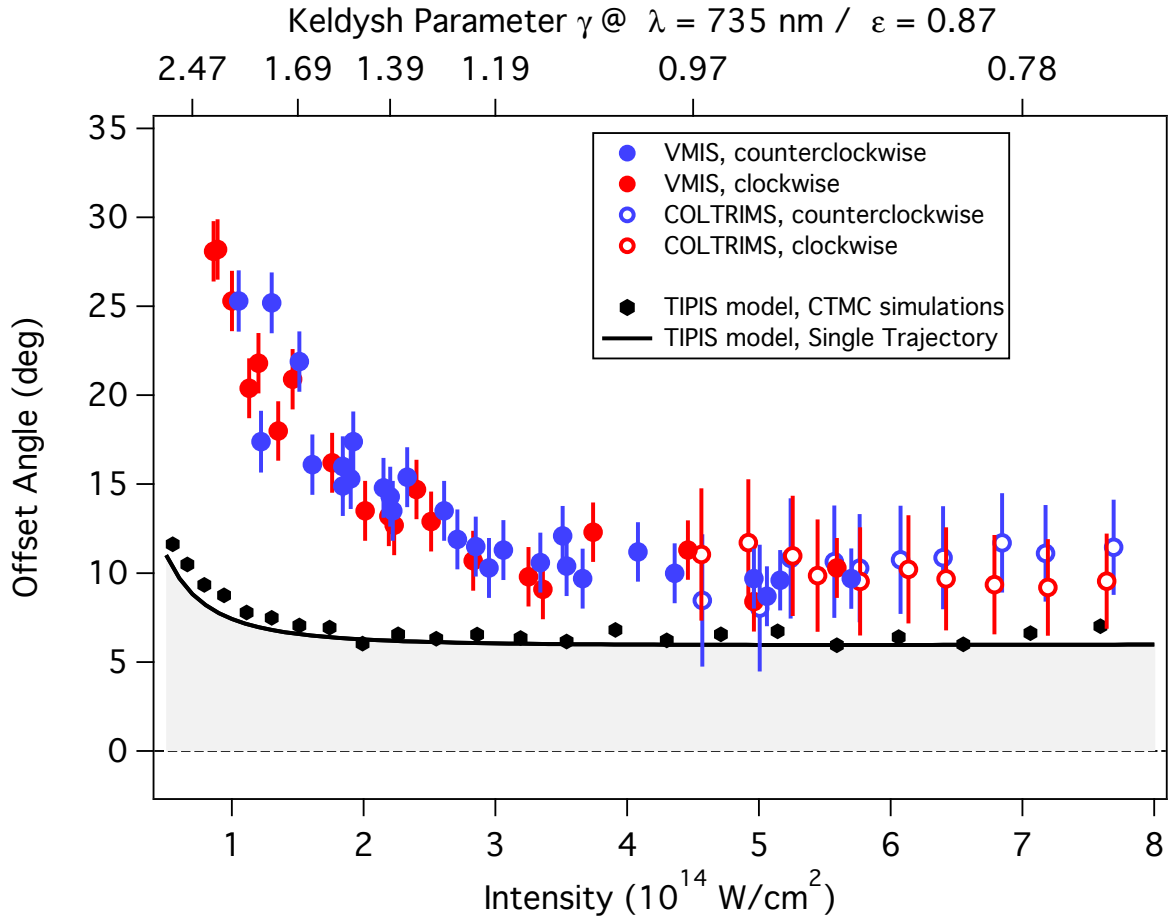

Fig. 2. Offset angle: $\theta_m - \pi/2$, from which tunneling time is extracted. Black line and dots correspond to the Coulomb correction, obtained using the TIPIS model [18] with single trajectory and classical trajectory Monte Carlo (CTMC) simulations, respectively (more in S.2).

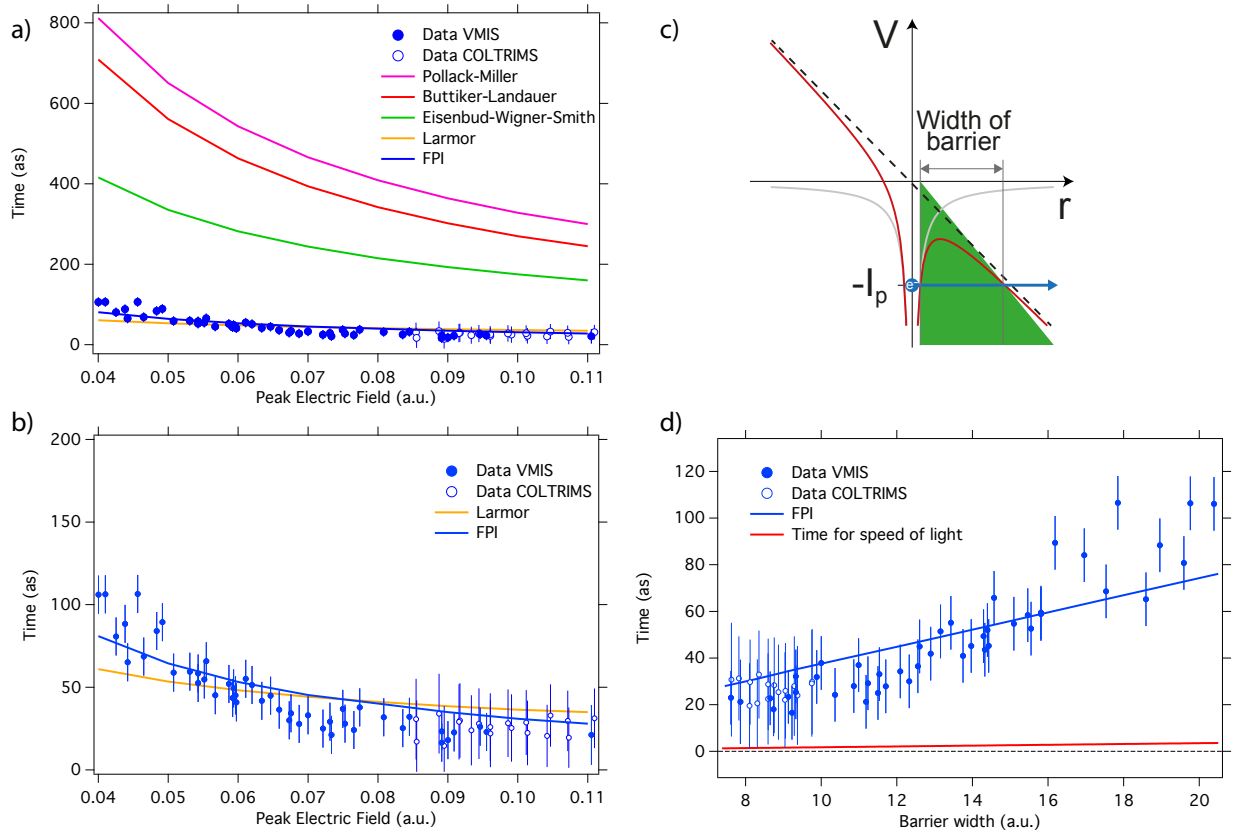

Fig. 3. Comparison of theory to experiment. a) Five theoretical predictions of tunneling time compared to experiment. b) Same as a), but zoomed in. c) Potential resulting from the combined Coulomb-Laser field. d) Tunneling time as a function of barrier width.

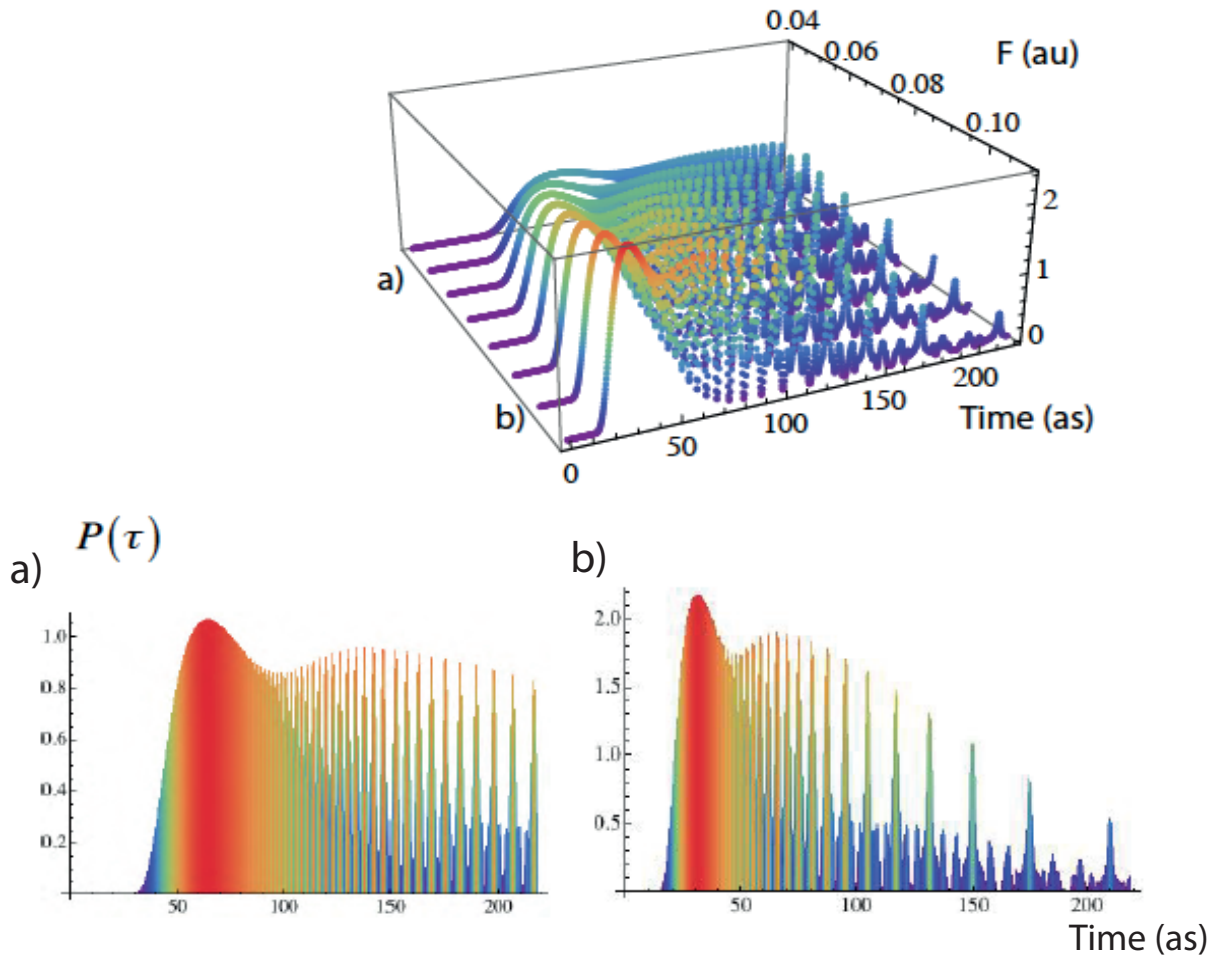

Fig. 4. A normalized distribution where each probability value corresponds to an integer number of attoseconds of tunneling time.  Top: 3D probability distribution of tunneling times for intensities ranging from 1 to $7.5 \times 10^{14} \, W/cm^2$ , corresponding to field strengths of 0.04 -0.11 au. The highly oscillatory structure is due to interference and shows tunneling probability peaking sharply at discrete values as tunneling time increases.  a) Probability distribution at Intensity = $1.625 \times 10^{14} W/cm^2$, FWHM $\approx$ 80 as, Skewness = 0.9.  b) Intensity = $6.5 \times 10^{14} \, W/cm^2$ , FWHM $\approx$ 50 as, Skewness = 1.09.